\documentclass[a4paper,11pt]{article}
\usepackage{pos}
\usepackage{lineno}

\title{Follow-up comments to searches for QCD instantons via forward proton tagging}

\author*[a]{Marek Ta\v{s}evsk\'y}
\author[b]{Valery Khoze}
\author[c]{Misha Ryskin}

\affiliation[a]{Institute of Physics of the Czech Academy of Sciences, Na Slovance 1999/2, 18221 Prague, Czech Republic}

\affiliation[b]{Department of Physics, University of Durham, DH1 3LE Durham, United Kingdom}

\affiliation[c]{Petersburg Nuclear Physics Institute, NRC ``Kurchatov Institute'', Gatchina, 188300 St. Petersburg, Russia}

\emailAdd{Marek.Tasevsky@cern.ch}
\emailAdd{v.a.khoze@durham.ac.uk}
\emailAdd{ryskin@thd.pnpi.spb.ru}

\abstract{In this short text, only the main facts are reminded from the
  full-fledged study~\cite{Tasevsky:2022sch} and a few experimental
  considerations are added, reflecting developments since the publication
  of Ref.~\cite{Tasevsky:2022sch} in 2023.}

\FullConference{The 33rd International Workshop on Deep Inelastic Scattering
  and Related Subjects (DIS2026)\\
4 - 8 May 2026\\
Bologna, Italy\\}

\begin{document}
\maketitle

\section{Considerations}
Although predicted by Quantum Chromodynamics, instantons evade their
experimental detection, mainly due to their complicated topological nature. The
instanton is not a particle but rather a family of objects of different sizes
and orientations in the Lorentz and color spaces. At the detector level, such
a final state is characterized by a high multiplicity of tracks with relatively
low momenta and a high sphericity. But since the bulk of so-called minimum bias
data at LHC shows the same properties and disposes of a huge cross section (the
main contribution coming from multi-parton interactions, MPI), it is evident
that any extraction of the instanton signal at LHC will be challenging and most
likely model-dependent. In general, data with zero or very low contamination by
pile-up need to be used since pile-up would lead to smearing any discriminating
variables and lowering various efficiencies. Recently, both ATLAS and CMS
pursued studies related to QCD instanton searches making use of minimum-bias
data with very low amounts of pile-up. 

In Ref.~\cite{CMS:2025sws} CMS have corrected data for detector effects and
provided distributions of event shape observables such as sphericity, thrust,
broadening and isotropy which are important in probing soft and nonperturbative
effects in QCD at the LHC such as e.g. quark-gluon plasma and QCD instantons.
A mismodeling observed when compared to predictions of numerous models
used in Ref.~\cite{CMS:2025sws} did not allow to consider the instanton
extraction but rather called for developing those models further which seems to
be critical for understanding the above phenomena. ATLAS analysis, at the moment
public in the form of PhD thesis~\cite{Ynyr_thesis}, delivers detector-level
blinded results making use of machine learning technique based on Deep Neural
Networks for preliminary signal discrimination and background estimation. The
data have so far been unblinded and the paper draft is undergoing a publication
process in the ATLAS collaboration. The biggest challenge in inclusive (or
minimum-bias) data analyses is the presence of the MPI background as explained
above. One of the ways to tackle this problem is to rely on diffractive events
characterized by large rapidity gaps and/or intact forward proton
tagging~\cite{Khoze:2021jkd}.
In our feasibility study~\cite{Tasevsky:2022sch}, we examined such a topology
and tried to account for all relevant effects with which experimental searches
will have to necessarily deal with. We included MPI into the background
estimates based on predictions from PYTHIA~8.2~\cite{Sjostrand:2014zea} as well
as detector and pile-up effects for signal and backgrounds, all using
DELPHES~3.5\cite{deFavereau:2013fsa} fast simulation package, leading not only
to deteriorating for example track efficiencies in the central detector but also
leading to a combinatorial background which stems from detecting a fake pile-up
proton in the forward proton detector AFP~\cite{Adamczyk:2015cjy,Tasevsky:2015xya} or CT-PPS~\cite{CMS:2014sdw}. After examining several cut scenarios,
the following ones (separately for $M_{\rm inst} > 60$~GeV and $M_{\rm inst} > 100$~GeV) give the best signal to background ratios (S/B):
\begin{equation}
  N_{\rm tr05} > 25 {\rm \ \ and \ \ } N_{\rm tr20} = 0 {\rm \ \ and \ \ } \sum E_T^{\rm fwcalo} < 5~{\rm GeV \ \ and \ 
\ }
  \xi^{\rm calo} < 0.025
  \label{Eq:goldendet60}
\end{equation}

\begin{equation}
N_{\rm tr05} > 30 {\rm \ \ and \ \ } N_{\rm tr25} = 0 {\rm \ \ and \ \ } \sum E_T^{\rm fwcalo} < 5~{\rm GeV \ \ and \ \ 
}
  \xi^{\rm calo} < 0.025
  \label{Eq:goldendet100}
\end{equation}

where $N_{\rm tr05}$, $N_{\rm tr25}$ and $N_{\rm tr20}$ are numbers of tracks in the region $0.0 < \eta < 2.0$
and for $p_T > 0.5$~GeV, $p_T > 2.5$~GeV and  $p_T > 2.0$~GeV, respectively, and $E_T^{\rm fwcalo}$ is a sum
of $E_T$ of clusters in the forward calorimeter with $p_T > 0.5$~GeV and $2.5 < \eta < 4.9$. The $\xi^{\rm calo}$
quantity is calculated as a sum of $E_{\rm T}e^{-\eta}/\sqrt{s}$ over calorimeter clusters with $E_{\rm T} > 0.2$~GeV (where $\sqrt{s} = 14$~TeV is the center-of-mass energy of colliding protons). 

\begin{figure*}
\includegraphics[width=0.5\textwidth,height=6cm]{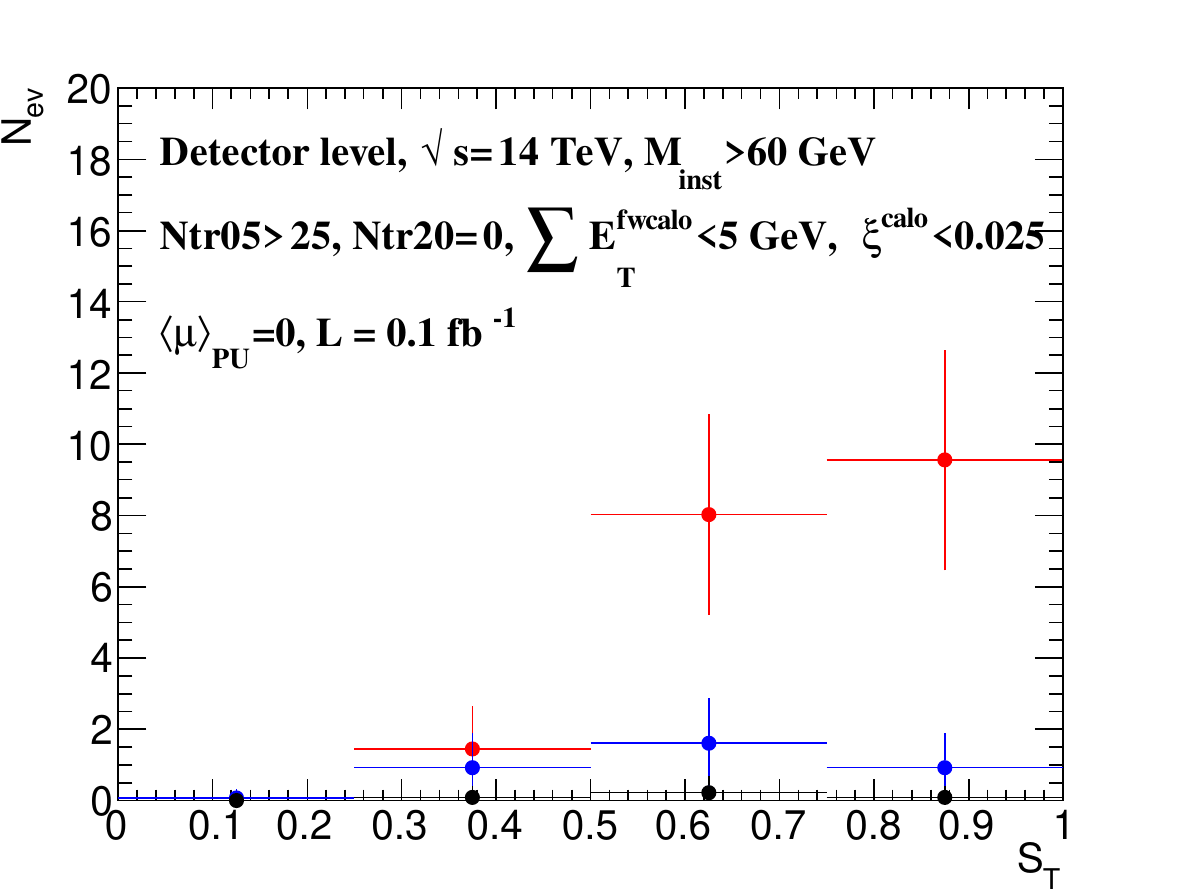}
\includegraphics[width=0.5\textwidth,height=6cm]{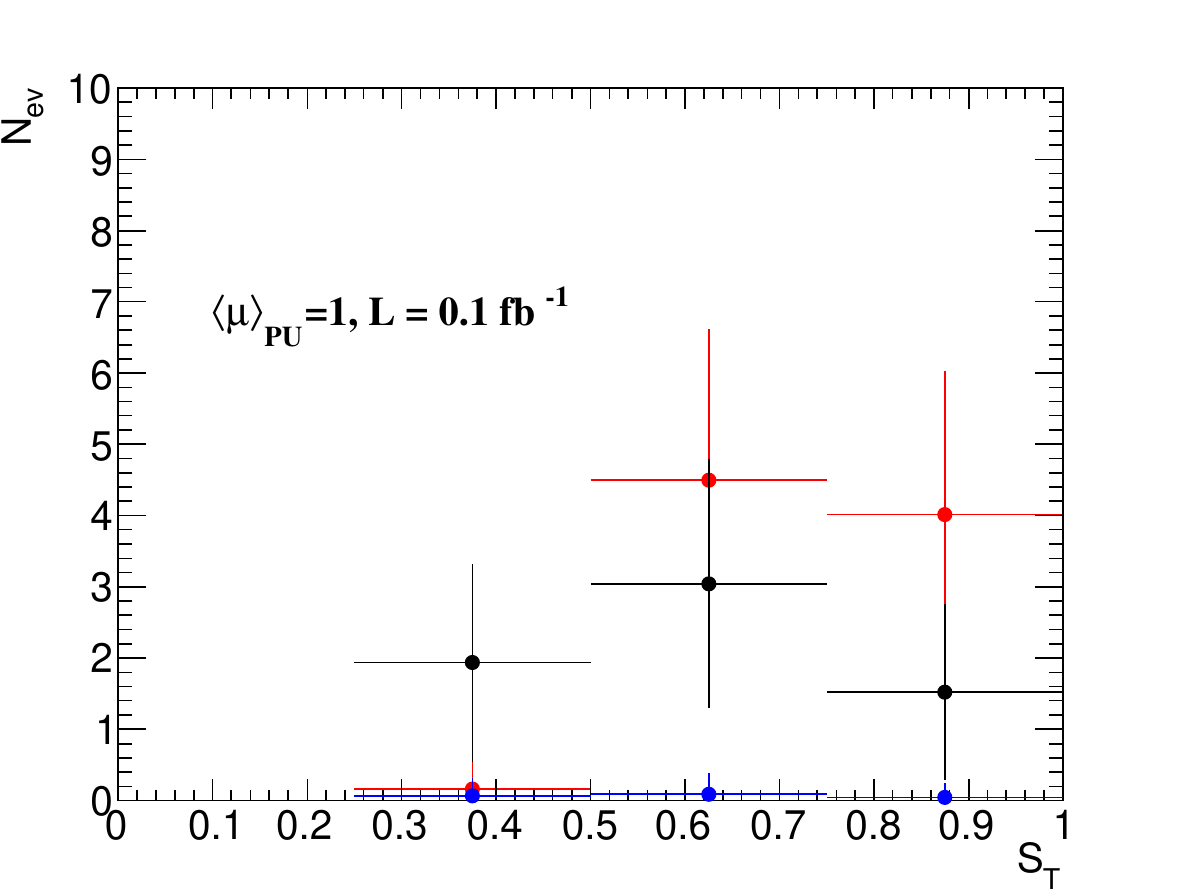}
\includegraphics[width=0.5\textwidth,height=6cm]{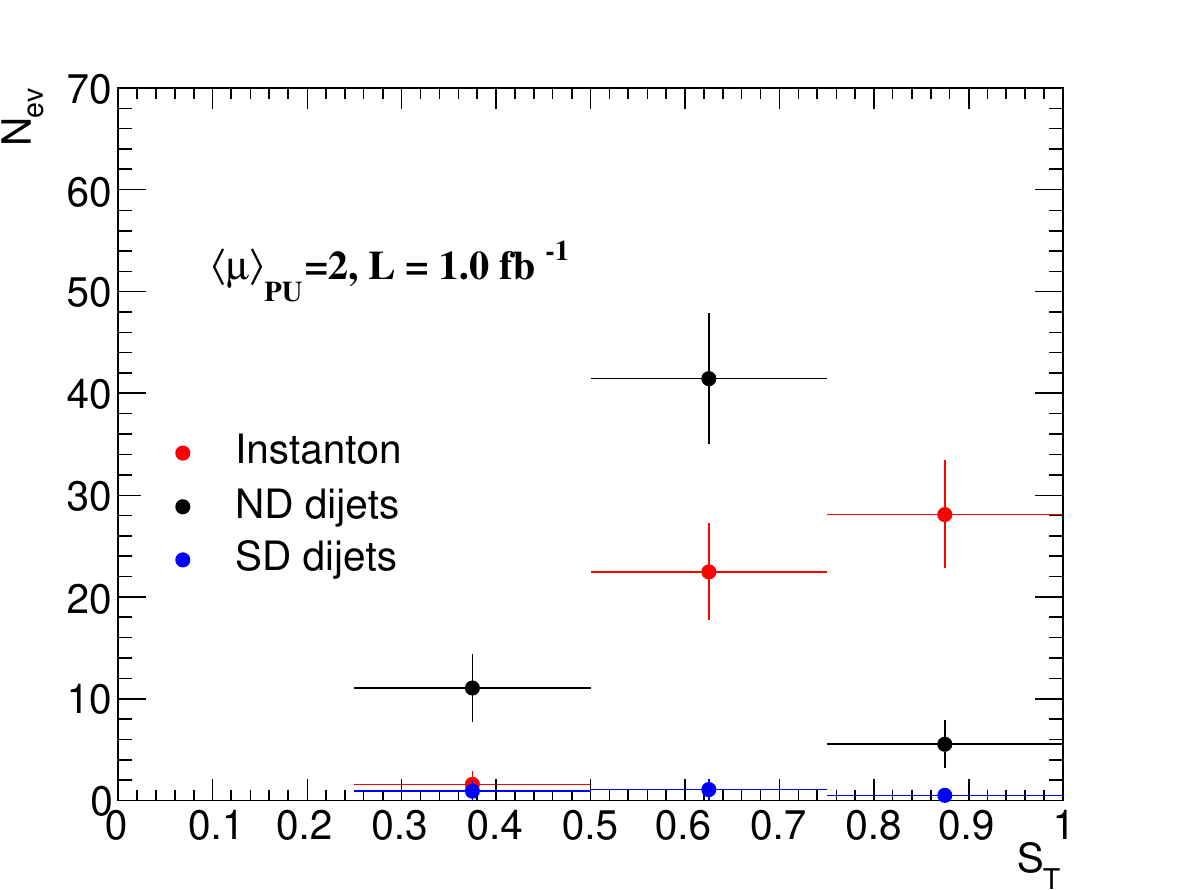}
\includegraphics[width=0.5\textwidth,height=6cm]{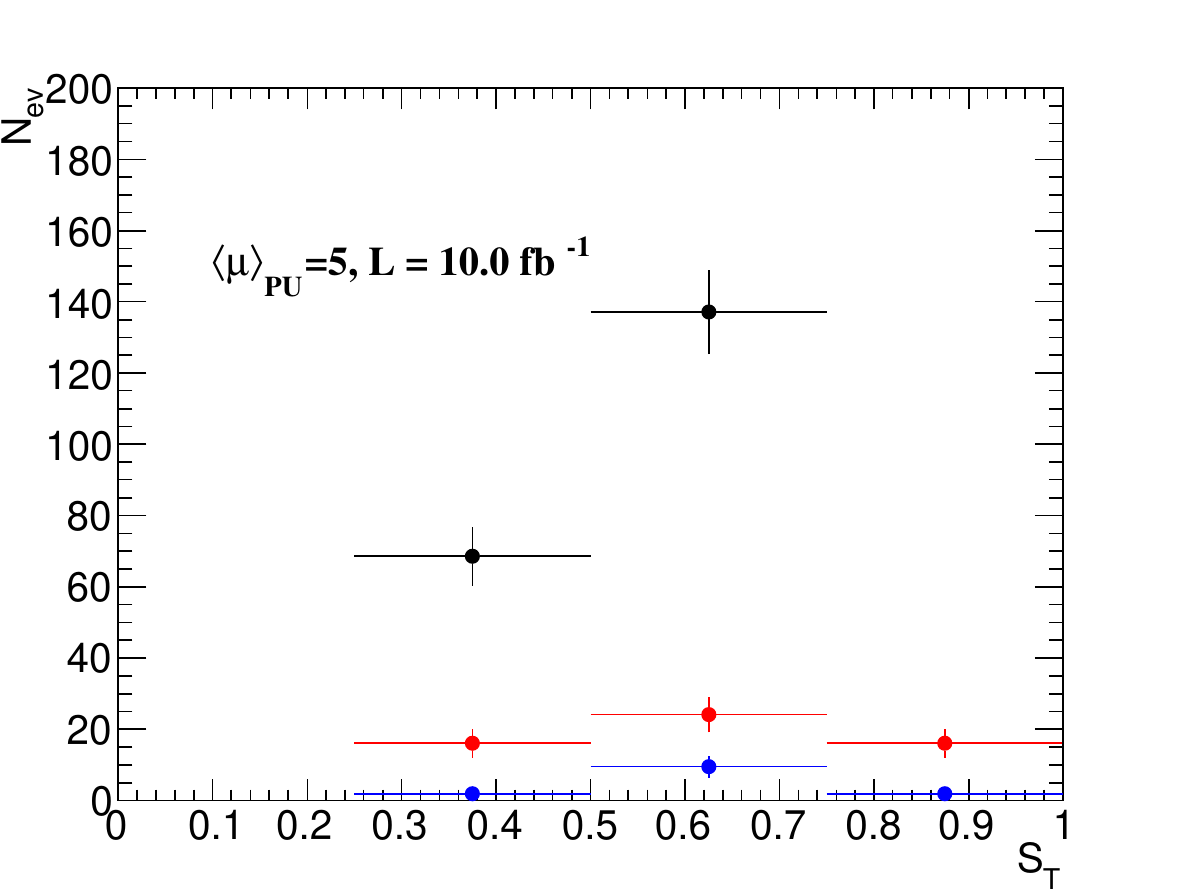}  
\caption{Distributions of expected event yields as functions of transverse
  sphericity at detector level for instanton signal from proton-Pomeron
  collisions generated by RAMBO for $M_{\rm inst} > 60$~GeV and backgrounds from
  ND dijets and SD dijets generated by PYTHIA~8.2 after applying detector-level
  cuts specified in Eq.~\ref{Eq:goldendet60} for four luminosity
  scenarios. Only statistical uncertainties are shown, estimated using expected
  event numbers from Table~1 in Ref.~[1]}
\label{fig:stdet60}
\end{figure*}

Four luminosity scenarios are considered with varying amounts of integrated
luminosity and pile-up, the latter characterized by an average number of pile-up
interactions per bunch crossing, $\mu$. To illustrate the situation at the
detector level for $M_{\rm inst} > 60$~GeV, in Figure~\ref{fig:stdet60} we show
distributions of transverse sphericity with expected numbers of events for all
luminosity scenarios (see also Table~1 in Ref.~\cite{Tasevsky:2022sch}). Results
in the first two scenarios, namely with $\mu=0$ and $\mu=1$, both with
integrated luminosity of 100~pb$^{-1}$, suggest that the instanton
signal should be separable from both combinatorial backgrounds, namely
Non-diffractive dijets and Single-diffractive dijets both overlayed with pile-up
in the central detector as well as in the forward proton detector.

As the LHC Run 3 has come to an end, it is possible to investigate how big data
samples depending on the amount of pile-up considered in
Ref.~\cite{Tasevsky:2022sch} are actually available. It is, however, necessary
to note that all considered pile-up amounts are only conceivable during the
so-called special low pile-up runs and as such, their durations and pile-up
profiles are a subject of negotiations inside the ATLAS and CMS collaborations
as well as of negotiations including the LHC management and all LHC experiments.
To our knowledge, the topic of searching for QCD instantons has not so far been
utilized as a part of any physics programmes used as motivations for collected
special runs. With the above limitations, rough estimates indicate data samples
corresponding to recorded integrated luminosities of around 1~pb$^{-1}$ with
$\mu <$~0.5 collected by ATLAS at energies of 13 or 13.6~TeV and about
4.7~pb$^{-1}$ with $0.1 < \mu < 0.3$ collected by CMS at 13~TeV. For higher
$\mu$, there are ATLAS samples of about 150~pb$^{-1}$ with $\mu=2$ at 13~TeV and
about 1.2~fb$^{-1}$ with $\mu=3$ at 13.6~TeV. CMS collected data samples of
about 300~pb$^{-1}$ with $\mu=2$ at 5.02~TeV, about 200~pb$^{-1}$ with $\mu=3$ at
13~TeV and about 2~fb$^{-1}$ with $\mu=5$ at 13.6~TeV. 

The above findings suggest that if the QCD instanton search would be restricted
to only data samples from Run~2 and Run~3 special runs, the separation of the
signal from backgrounds would worsen compared to the situation presented in
Fig.~\ref{fig:stdet60} owing to increased statistical uncertainties. There are,
however, improvements which should not cost too much, as elaborated in
Ref.~\cite{Tasevsky:2022sch}. With existing data, one can utilize an
event-by-event maximum particle density and the fact that in signal, more
strange and charm particles are produced than in the backgrounds. If special
runs can be negotiated also in Run~4, in addition to the above,
a dedicated instanton trigger and a timing information from the central tracker
would certainly improve the S/B ratio. 

\acknowledgments
The authors are grateful to Michael Pitt for providing integrating luminosities
of special runs collected by CMS.

\end{document}